\documentclass[pre,twocolumn,superscriptaddress,showpacs]{revtex4}

\ifx\pdftexversion\undefined
  \usepackage[dvips]{graphics}
  \DeclareGraphicsExtensions{.eps}
\else
  \usepackage[pdftex]{graphics}
  \DeclareGraphicsExtensions{.pdf}
\fi

\usepackage{amsfonts}
\usepackage{amsmath}
\usepackage{amssymb}

\begin{document}
\title{Dephasing representation: Employing the shadowing theorem to calculate quantum correlation functions}
\author{Ji\v{r}\'{\i} Van\'{\i}\v{c}ek}
\email{vanicek@post.harvard.edu}
\affiliation{Mathematical Sciences Research Institute, Berkeley, California 94720, USA}
\affiliation{Department of Chemistry, University of California, Berkeley, California 94720, USA}
\date{2 April 2004}

\begin{abstract}
Due to the Heisenberg uncertainty principle, various classical systems differing only on the scale
smaller than Planck's cell correspond to the same quantum system. This fact is used to find a unique semiclassical representation
without the Van Vleck determinant, applicable to a large class of correlation functions expressible as quantum fidelity. As in the
Feynman path integral formulation of quantum mechanics, all contributing
trajectories  have the same amplitude: that is why it is
denoted the ``dephasing representation.''  By
relating the present approach to the problem of existence of true
trajectories near numerically-computed chaotic trajectories, the
approximation is made rigorous for any system in which the shadowing
theorem holds. Numerical implementation only requires computing actions along
the unperturbed trajectories and not finding the shadowing trajectories. While semiclassical linear-response theory was used before in quasi-integrable and chaotic systems, here its validity is justified in the most generic, mixed systems. 
Dephasing representation appears to be a rare practical method to calculate quantum correlation functions in nonuniversal regimes in many-dimensional systems where exact quantum calculations are impossible.  
\end{abstract}

\maketitle

The method that is described in this paper is based on two observations:
the first is the fact that\ the relationship between classical and quantum
dynamics is many to one; the second is the idea of shadowing of a perturbed
trajectory by a nearby unperturbed trajectory. We start by explaining these
two ingredients in more detail.

\emph{Classical vs quantum dynamics}. In the semiclassical (SC) approximation,
quantum wave function is associated with a classical surface (Lagrangian
manifold) in phase space. SC evolution of the wave function is
performed by classically evolving this surface and computing actions
along the trajectories of the points of the surface. At the end, the surface
is projected onto an appropriate coordinate plane. If we slightly distort
the initial surface, individual trajectories will change exponentially
fast. However, if the distortion is small enough, the original and distorted
initial surfaces semiclassically correspond to the same quantum
wave functions (their overlap is $\approx 1$). Due to the unitarity of quantum evolution, the overlap of the two
wave functions associated with the two evolved surfaces will remain $\approx 1$ for all times.

\emph{Shadowing}. Because of the exponential sensitivity to initial
conditions and because of the finite precision of a computer,
computer generated trajectories in chaotic systems are accurate only for a
logarithmically short time. As a result, it was not clear whether it makes sense at all to do computer simulations for longer times and whether, e.g., the fractal patterns seen in these simulations are real. The solution was offered by Hammel,
Yorke, and Grebogi \cite{HAMMEL}  with the
concept of shadowing which was later, in various settings, promoted to a
theorem \cite{GREBOGI,SHADOWING}. Their finding is that while a  computed trajectory
diverges exponentially from the true trajectory with the same initial
conditions, there exists an errorless trajectory with a slightly different
initial condition that stays near (``shadows'') the  computed one.

While shadowing is extremely useful if the perturbation is random, which is
the case for the roundoff computer errors, it works also if the perturbation
is deterministic, e.g., given by a precise change of the Hamiltonian. This
is a completely different physical problem: Even if a computer could evolve
trajectories exactly without any roundoff errors, we might be interested in
comparing the trajectories with the same initial conditions, but of two slightly
different Hamiltonians. It turns out that in chaotic systems these
trajectories will again exponentially diverge. Below we rigorously show that
the latter problem can be transformed to the former.

Comparison of the dynamics of two slightly different Hamiltonians is the
subject of a very rich and recently much studied subject of the sensitivity
of classical and quantum motion to perturbations \cite{PERES,JALABERT,CERRUTI,FIDELITY,VANICEK,FIDELITY_PAPERS}. This sensitivity is best
described by so-called fidelity $M\left( t\right) $: Classical (quantum)
fidelity is the classical (quantum) overlap at time $t$ of two initially
identical phase-space distributions (quantum states $|\psi \rangle $), that
were evolved by slightly different Hamiltonians $H^{0}$ and $H^{\epsilon
}=H^{0}+\epsilon V$ where $\epsilon $ controls the strength of the
perturbation. In Dirac notation, %
\[
M\left( t\right) = \left\vert O(t) \right\vert^2 = \left\vert \left\langle \psi \left\vert
e^{+i(H^{0}+\epsilon V)t/\hbar }e^{-iH^{0}t/\hbar }\right\vert \psi
\right\rangle \right\vert ^{2}.
\]

Studying fidelity is useful in itself but also because many other, seemingly unrelated quantum correlation functions take the same mathematical form. For example, $M(t)$ is equivalent to the Loschmidt echo---the survival probability of a state evolved first by $H^0$ for time $t$, then by $-H^{\epsilon}$ for time $t$. If we were to evaluate this overlap directly semiclassically \cite{JALABERT,FIDELITY_PAPERS,CERRUTI}, we would have to find
exact trajectories of $H^{0}$ and of $H^{\epsilon }$, corresponding phases
and prefactors, and then calculate the overlap. This would be extremely
difficult in chaotic systems because of the exponential growth of errors,
exponentially proliferating contributing trajectories, and exponentially
growing number of singularities. In fact, full SC treatment, i.e., evaluating the overlap \emph{itself} by the stationary phase approximation, would yield anything but the right answer since we would be adding up a huge number of infinite singularities.  Compared to the direct approach, the dephasing representation (DR) is simple, as can be seen from the following alternative expressions:
\begin{align} 
\label{dephasing_representation}
O_{DR}(t) &=  \left(
2\pi \hbar \right) ^{-d} \int d^{d}p^{\prime }\exp \left[ i
\Delta S\left( \mathbf{r}^{\prime },\mathbf{p}^{\prime },t\right)/\hbar
\right],\\
O_{DR}(t) &= \int d^{d}r^{\prime }\exp \left[i \Delta S\left( 
\mathbf{r}^{\prime},\mathbf{p}^{\prime },t\right) /\hbar \right] \left\vert \psi \left( 
\mathbf{r}^{\prime }\right) \right\vert ^{2},  \nonumber \\
O_{DR}(t) &= \int d^{d}r^{\prime }\int d^{d}p^{\prime }\exp 
\left[ \frac{i}{\hbar }\Delta S\left( \mathbf{r}^{\prime },\mathbf{p}%
^{\prime },t\right) \right] \rho_W\left( \mathbf{r}^{\prime },\mathbf{p}^{\prime
}\right), \nonumber
\end{align}
where the first expression is for position eigenstates $|\mathbf{r}^{\prime }\rangle $, the second for wavepackets with mean momentum $\mathbf{p}^{\prime
}$, and the third (and most general) for Wigner distributions $\rho_{W}$. Since Wigner distribution can represent mixed states, the third expression can describe purity fidelity and decoherence. Analogs of the first two expressions exist in momentum and other representations, and all can be derived from the Wigner form. In all expressions, the action difference $\Delta S$ is just the integral of the perturbation along the unperturbed trajectory with initial conditions $\mathbf{r}^{\prime }$, $%
\mathbf{p}^{\prime }$, 
\[
\Delta S\left( \mathbf{r}^{\prime },\mathbf{p}^{\prime },t\right)
=S^{\epsilon}-S^{0}=- \epsilon \int_{0}^{t}d\tau \,V\left[ \mathbf{r}\left( \tau \right)
,\tau \right]. 
\]
 Roughly speaking, the motivation for DR is the following: instead of using the
trajectory of $H^{\epsilon }$ with the same initial condition, we use the shadowing trajectory: a trajectory of $H^{\epsilon }$ with a slightly different initial condition, which remains close to the trajectory 
of $H^{0}$ up to time $t$. [That this is possible follows from the shadowing theorem (ST) as we show below.] All we have to compute is the phase difference along the
trajectory. We avoid the singularities since the prefactors precisely
cancel and we also need to evaluate a much smaller number of trajectories
than in the standard SC treatment. We explain these claims below when we derive the approximation.

First let us show that the problem of sensitivity of dynamics to the change
of Hamiltonian is equivalent to the problem of shadowing of numerically
noisy trajectories. For simplicity we show this for a general
two-dimensional symplectic map, which is nothing else but a discretization
of continuous-time dynamics of a Hamiltonian system. We need three
standard definitions that hold for any (not-necessarily symplectic) map $f:%
\mathbf{x}_{n}\mapsto \mathbf{x}_{n+1}$.

\textbf{Definition}. A \emph{true trajectory} $\left\{ \mathbf{x}%
_{n}\right\} _{n=a}^{b}$ of a map $f$ satisfies $\mathbf{x}_{n+1}=f(\mathbf{x%
}_{n})$ for $a\leq n\leq b$.

\textbf{Definition}. $\left\{ \mathbf{\tilde{x}}_{n}\right\} _{n=a}^{b}$ is
an $\epsilon$-\emph{pseudotrajectory} of a map $f$ if $\left\vert 
\mathbf{\tilde{x}}_{n+1}-f(\mathbf{\tilde{x}}_{n})\right\vert <\epsilon $
for $a\leq n\leq b$.

\textbf{Definition}. The true trajectory $\left\{ \mathbf{x}_{n}\right\}
_{n=a}^{b}$ $\delta $-\emph{shadows} $\left\{ \mathbf{\tilde{x}}_{n}\right\}
_{n=a}^{b}$ on $a\leq n\leq b$ if $\left\vert \mathbf{x}_{n}-\mathbf{\tilde{x%
}}_{n}\right\vert <\delta $ for $a\leq n\leq b$.

Anosov \cite{ANOSOV} and Bowen \cite{BOWEN} showed that for uniformly hyperbolic maps, noisy trajectories
can be shadowed for an infinitely long time by true trajectories. Most
physical systems, however, are not uniformly hyperbolic, which led to a
series of investigations for more general systems. By now, there exist theorems for both continuous differential equations and discrete
maps, and they require the satisfaction of various assumptions \cite{HAMMEL,GREBOGI,SHADOWING}. The general conclusion is that shadowing
in nonuniformly hyperbolic systems only works for a finite time $t_S$, which decreases with
increasing perturbation. Below we relate the problem of Hamiltonian
perturbations to the problem of the shadowing of noisy trajectories for
symplectic maps, without considering detailed assumptions of a particular ST.

Let $q_{n}$ and $p_{n}$ be the position and momentum at time $n$ and
let $\mathbf{x}_{n}=\left( q_{n},p_{n}\right) $. Let the continuous system
be described by a Hamiltonian $H^{\epsilon }=H^{0}+\epsilon
V=p^{2}/2+W +\epsilon V$, where $W$ is the unperturbed
potential and $\epsilon V$ is the perturbation. Further assume that $V$ and $%
W$ are differentiable and $\left\vert \nabla V \left( q, t \right)
\right\vert <1$ for all $q$. The corresponding symplectic map $f^{\epsilon }:%
\mathbf{x}_{n}\mapsto \mathbf{x}_{n+1}$ is given by%
\begin{eqnarray*}
p_{n+1} &=&p_{n}-\nabla W \left( q_{n}, n \right) -\epsilon \nabla V \left(
q_{n}, n\right) , \\
q_{n+1} &=&q_{n}+p_{n+1}.
\end{eqnarray*}

\textbf{Lemma 1}. The true trajectory $\left\{ \mathbf{\tilde{x}}_{n}\right\}
_{n=a}^{b}$ of $f^{\epsilon }$ is an $\epsilon $-pseudotrajectory of $f^{0}$.

{\it Proof.} $\left\vert \mathbf{\tilde{x}}_{n+1}-f^{0}(\mathbf{\tilde{x}}%
_{n})\right\vert =\left\vert f^{\epsilon }(\mathbf{\tilde{x}}_{n})-f^{0}(%
\mathbf{\tilde{x}}_{n})\right\vert =\left\vert \epsilon V^{\prime }\left( 
\tilde{q}_{n}\right) \right\vert <\epsilon $.

\textbf{Lemma 2}. Let the noise amplitude be $\delta $, i.e., let the
numerically computed trajectory $\left\{ \mathbf{\bar{x}}_{n}\right\}
_{n=a}^{b}$ of $f^{\epsilon }$ be a $\delta $-pseudotrajectory of $%
f^{\epsilon }$. Then $\left\{ \mathbf{\bar{x}}_{n}\right\} _{n=a}^{b}$ is a $%
\left( \delta +\epsilon \right) $-pseudotrajectory of $f^{0}$.

{\it Proof.} $\left\vert \mathbf{\bar{x}}_{n+1}-f^{0}(\mathbf{\bar{x}}%
_{n})\right\vert =\left\vert \mathbf{\bar{x}}_{n+1}-f^{\epsilon }(\mathbf{%
\bar{x}}_{n})+f^{\epsilon }(\mathbf{\bar{x}}_{n})-f^{0}(\mathbf{\bar{x}}%
_{n})\right\vert <\left\vert \mathbf{\bar{x}}_{n+1}-f^{\epsilon }(\mathbf{%
\bar{x}}_{n})\right\vert +\left\vert f^{\epsilon }(\mathbf{\bar{x}}%
_{n})-f^{0}(\mathbf{\bar{x}}_{n})\right\vert <\delta +\left\vert f^{\epsilon
}(\mathbf{\bar{x}}_{n})-f^{0}(\mathbf{\bar{x}}_{n})\right\vert <\delta
+\epsilon $, where we used Lemma 1 in the last inequality.

Lemma 1 shows that Hamiltonian perturbations are equivalent to random noise
as long as further assumptions of the particular ST are
satisfied. Lemma 2 shows that we can combine Hamiltonian perturbations with
random noise before applying the ST.

The author believes that the DR expressions (\ref{dephasing_representation}) are the fundamental representations of fidelity because they provide (to his knowledge) not only the only accurate but simply the only way to calculate fidelity semiclassically for longer than the logarithmic time. DR can be thought of as the Van Vleck propagator in the mixed position-momentum representation, so that the Van Vleck determinant equals $\det (\partial \mathbf{p}_{\mathrm{final}} / \partial \mathbf{p}^{\prime}) = 1$ and the Maslov indices cancel. However, since most authors are familiar only with the Van Vleck propagator in the position representation, we show explicitly how DR can be derived by a \emph{careful} treatment of this propagator.

The Van Vleck propagator in position representation,%
\[
K_{SC}(\mathbf{r}^{\prime \prime },\mathbf{r}^{\prime };t)=\sum_{j}\left(
2\pi i\hbar \right) ^{-d/2}C_{j}^{1/2}\exp \left( \frac{i}{\hbar }S_{j}-%
\frac{i\pi }{2}\nu _{j}\right) ,
\]
is the standard SC approximation of the quantum propagator. 
The sum is over all classical trajectories connecting point $\mathbf{r}%
^{\prime }$ at time $\tau=0$ to point $\mathbf{r}^{\prime \prime }$ at time $\tau =t$, $S_{j}$ is the classical action,%
\[
S_{j}\left( \mathbf{r}^{\prime \prime },\mathbf{r}^{\prime };t\right)
=\int_{0}^{t}d\tau\,L\left[ \mathbf{r}\left( \tau\right) \mathbf{%
,\dot{r}}\left( \tau \right), \tau \right] ,
\]%
and $C_{j}=\vert \det ( \partial ^{2}S_{j}/\partial \mathbf{%
r}^{\prime \prime }\partial \mathbf{r}^{\prime } ) \vert $ is
the Van Vleck determinant equal to the classical transition
probability. There are three major problems with this approximation. First,
the number of trajectories in a generic system grows exponentially
with time. Second, there is a growing number of singularities in this
expression at conjugate points. Third, direct evaluation of this
expression would involve an expensive \emph{root search}, i.e.,
finding for each trajectory $j$ the initial momentum $\mathbf{p}_{j}^{\prime }
$ leading to the final position $\mathbf{r}^{\prime \prime }$. In improved
SC methods such as the initial value representation, the root search
and singularities are avoided, at a cost of replacing the sum over 
trajectories $j$ by an integral over initial conditions 
\cite{MILLER_IVR, MILLER_IVR_REVIEW}.

Forgetting \emph{for a moment} all these problems (as have forgotten most authors discussing fidelity semiclassically \cite{JALABERT,FIDELITY_PAPERS}, with the exception of Ref.~\cite{CERRUTI} where fidelity is computed semiclassically for a logarithmically short time), let us use $K_{SC}$ to express fidelity amplitude:
\[
O(t) \approx \int d%
\mathbf{r} K_{SC}^{\epsilon \ast }\left( \mathbf{r},\mathbf{r}^{\prime };t\right) K_{SC}^{0}\left( \mathbf{r},\mathbf{r}^{\prime };t\right)
\]%
(The superscript tells us which Hamiltonian is used.) We could proceed in several ways: we could try evaluating this integral by the stationary-phase method. That would give a singular result as explained in the text above Eq.~(\ref{dephasing_representation}) because there can be many coalescing stationary phase points. (For disbelievers the author recommends trying it.) Or we could evaluate the integral numerically, which might smooth out some of the singularities in the two integrands by integration, but would be virtually impossible in practice. What Jalabert and Pastawski \cite{JALABERT} suggested was using only the diagonal terms ($j^0=j^{\epsilon}$), justifying this by the claim that the off-diagonal terms cancel out in the average over initial states (or realizations of the perturbation, since they calculated the average fidelity). There are several problems with this approach. Since $H^0$ and $H^{\epsilon}$ are different, after a long enough time it will not be possible to distinguish terms as diagonal and off-diagonal. Also it is not obvious that the off-diagonal terms should cancel since these are in majority.

However, this separation \emph{is} possible in DR. For the trajectories of $H^{\epsilon}$, instead of using the precise initial conditions, we infinitesimally adjust the initial conditions so that the trajectories of $H^{\epsilon}$ shadow those of $H^0$. While $K^0_{DR}=K^0_{SC}$ remains unchanged in DR,
\[
K^{\epsilon}_{DR}(\mathbf{r},\mathbf{r}^{\prime };t)=\sum_{j}\left(
2\pi i\hbar \right) ^{-d/2}(C_{j}^0)^{1/2}\exp \left( \frac{i}{\hbar }S^{\epsilon}_{j}-%
\frac{i\pi }{2}\nu^0 _{j}\right).
\]
Note that because we are using the shadowing trajectory, the Maslov index and the Van Vleck determinant are the same as for $K^0_{DR}$, only the action is different (since this trajectory \emph{is} a true trajectory of $H^{\epsilon}$). Now there is an exact one-to-one correspondence between the terms of the two propagators. Moreover, if we use the Loschmidt echo picture, only the diagonal terms form a continuous trajectory in phase space (momentum at time $t$ is continuous), and so semiclassically, only these terms survive. It turns out, that in DR, neglecting the off-diagonal terms actually improves the approximation. This can be seen, e.g., for $\epsilon=0$, where we get $O_{DR}(t)=1$ only if we neglect the off-diagonal terms (this was noted already by Jalabert and Pastawski \cite{JALABERT}). As a result, we have    
\[
O_{DR}(t) = \int d^{d}r \sum_{j}\left(
2\pi \hbar \right) ^{-d}C^0_{j}\exp \left( i \Delta S_{j} / \hbar\right).
\]
Realizing that $C^0_j = |\det (\partial \mathbf{p}^{\prime}/\partial \mathbf{r})|$, we obtain DR (\ref{dephasing_representation}) for position eigenstates $\vert \mathbf{r}^{\prime}\rangle$.
\begin{figure}[htbp]
\centerline{\resizebox{\hsize}{!}{\includegraphics{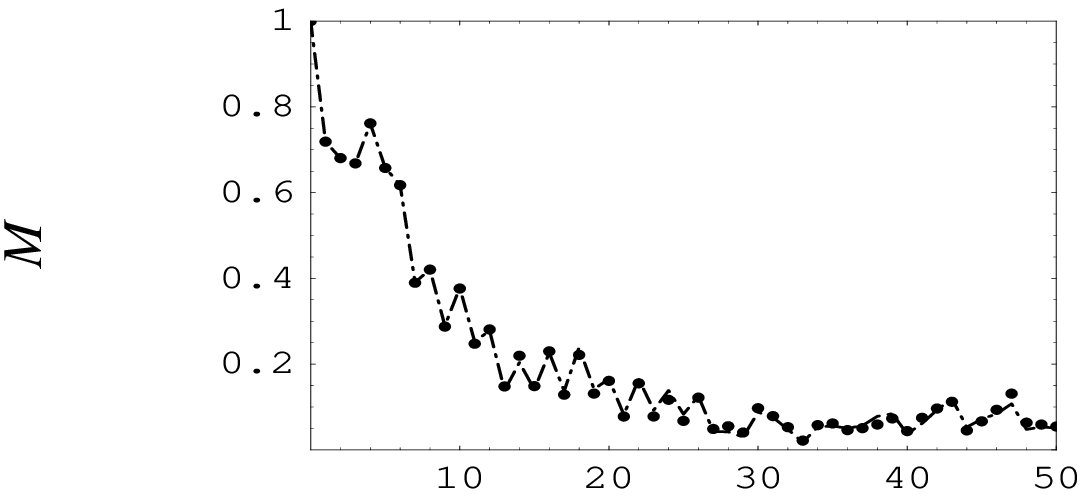}}}
\centerline{\resizebox{\hsize}{!}{\includegraphics{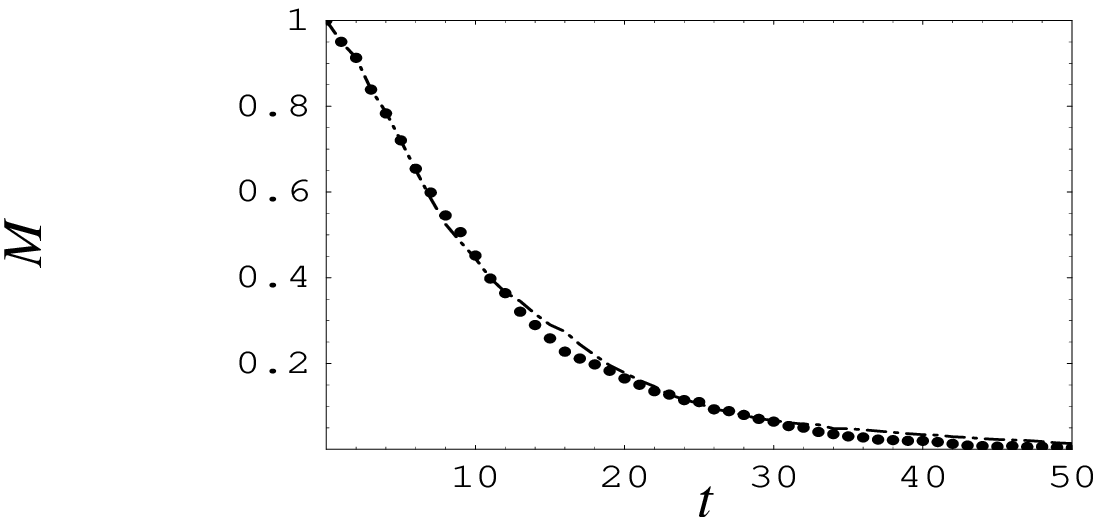}}}
\caption{\label{figure} Comparison of the exact quantum calculation (dots) and dephasing representation (dashed-dotted line) of fidelity $M(t)$ for the standard map ($n = 1000$, $1000$ classical trajectories used). Top: mixed phase space ($k=0.8$, $\epsilon = 5 \times 10^{-3}$). Bottom: chaotic phase space ($k=10$, $\epsilon = 2 \times 10^{-3}$).}
\end{figure}

DR was first numerically tested in chaotic systems which are ``closer'' to being uniformly hyperbolic \cite{FIDELITY,VANICEK}. Motivated by the ST for nonhyperbolic systems here we also apply the method to mixed systems. 
Figure~\ref{figure} shows fidelity decay of initial position states ($q=0.4$) for the Chirikov standard map ($W = kV= - k \cos q$, $\hbar = 1/2\pi n$) in cases with mixed and chaotic dynamics. The success of DR is striking: using only $1000$ trajectories in the chaotic case (bottom), the approximation is excellent at time $t=50$ where there are $\sim 10^{20}$ SC contributions in the standard approach \cite{JALABERT,FIDELITY_PAPERS,CERRUTI}. DR works since for chaotic systems, only the statistics of actions matters, which can be reproduced by a smaller ensemble \cite{FIDELITY,VANICEK}. Counterintuitively, experts in the field of chaos might be more surprised by the results for mixed phase space (top). The reason is that in both integrable \emph{and} chaotic systems there exist \emph{different} simplifications. In mixed systems with both invariant tori and chaotic regions, neither simplification works \emph{throughout} phase space. DR still reproduces even the fine details of $M(t)$ in Fig.~\ref{figure}. 

While there is a rigorous prescription how to verify the applicability of the ST, it is beyond the goal of this paper. It is done by numerically finding, for each trajectory, rigorous bounds on several dynamical quantities \cite{SHADOWING}. The goal of this paper was to reveal the surprising connection between classical shadowing and quantum mechanics and to present a generally applicable SC method. We therefore omit here the rigorous check of the assumptions of the ST for the specific example in Fig.~\ref{figure}. A rough estimate $t_S \sim \epsilon^{-1/2}$ for the shadowing time from a conjecture in Refs.~\cite{HAMMEL,GREBOGI,SHADOWING} gives $t_S \sim 14$ and $23$ for the top and bottom parts of Fig.~\ref{figure}, respectively. This estimate should be used with caution since it does not even depend on $k$. Figure~\ref{figure} suggests that most trajectories are shadowable well beyond this time, because otherwise it would be virtually impossible that a method based on the ST, using a large number of quite randomly interfering waves should mimic so accurately the exact quantum solution. 

A more thorough description of the universal numerical success of DR in the Gaussian, algebraic, Fermi-Golden-Rule, Lyapunov, and intermediate regimes, and for more general perturbations, will be presented elsewhere \cite{CONFERENCE}. Furthermore, it can be shown that considering statistics of action differences in DR leads to a simple unified theory of the four regimes \cite{CONFERENCE,UNIFIED}.
It should be noted that before the work in Ref.~\cite{FIDELITY}, due to the exponential proliferation of
SC contributions, numerical
evaluation of $M(t)$ for longer times was usually only done using exact quantum propagation on a grid \cite{FIDELITY_PAPERS}.  For the standard map that we used, the most
successful SC calculations were done up to time $t<10$ where
there were fewer than $10^{4}$ contributing classical trajectories 
\cite{CERRUTI}. In many-dimensional systems, DR seems to be the \emph{only} method to evaluate fidelity because the complexity of exact quantum propagation grows exponentially with dimensionality.

To conclude, DR is an efficient method that can describe quantum-mechanical decays purely in terms of dephasing. This strangely suggests that in some cases it is possible to describe the decay of classical overlaps by interference \cite{CONFERENCE}. In particular, we have rigorously shown that DR is accurate for systems and times where ST applies. 
The many-to-one relationship between  classical and quantum dynamics on which DR  is based has been exploited before \cite{KAPLAN,VANICEK1}. While the replacement-manifold method \cite{VANICEK1} is very
simple and gives excellent results in a variety of problems, it requires
finding the replacement manifolds. DR goes one step further: here the fact that a slightly distorted
initial surface exists is sufficient. It is not necessary to find it
explicitly.

The author conjectures that DR should be applicable to other classes of
correlation functions of the type $\langle \psi \vert e^{A^{\dag
}}e^{B}\vert \psi \rangle $ where operator $A$ differs only slightly from $B$, and $A$ and $B$ do not necessarily correspond to
real-time quantum evolution. Possible applications may include temporal and thermal correlation functions in
many-body systems in condensed matter and chemical physics, and in the general wave scattering in disordered media.

The author would like to thank E.J. Heller and W.H. Miller for discussions and the Mathematical Sciences Research Institute for support.


\begin{thebibliography}{99}  

\bibitem{HAMMEL} S.M. Hammel, J.A. Yorke, and C. Grebogi, J. Complex. {\bf 3}, 136 (1987).

\bibitem{GREBOGI} C. Grebogi {\it et al.}, Phys. Rev. Lett. {\bf 65}, 1527 
	(1990).

\bibitem{SHADOWING} 
	T. Sauer and J.A. Yorke, Nonlinearity {\bf 4}, 961 (1991); 
	S.-N. Chow and K.J. Palmer, J. Complex. {\bf 8}, 398 
	(1992);
	B.A. Coomes, H. Kocak, and K.J. Palmer, J. Comput. Appl. Math. {\bf 52}, 35 (1994); T. Sauer, C. Grebogi, and J. A. Yorke, Phys. Rev. Lett. {\bf 79}, 59 (1997);
	H. Kantz {\it et al.}, Phys. Rev. E {\bf 65}, 026209 (2002);
  	W. Hayes and K.R. Jackson, SIAM (Soc. Ind. Appl. Math.) J. Numer. Anal. {\bf 41}, 1948 (2003). 
\bibitem{PERES}
	A. Peres, Phys. Rev. A {\bf 30}, 1610 (1984). 


\bibitem{JALABERT} R.A. Jalabert and H.M. Pastawski, Phys. Rev. Lett. {\bf 86}, 2490 (2001).

\bibitem{FIDELITY} J. Van\'{\i}\v{c}ek and E.J. Heller, Physical Review E, {\bf 68}, 056208 (2003).

\bibitem{VANICEK} J. Van\'{\i}\v{c}ek, Ph.D. thesis, Harvard University, 2003, {\tt http://physics.harvard.edu/Thesespdfs/vanicek.pdf}


\bibitem{FIDELITY_PAPERS} 
	P. Jacquod, P.G. Silvestrov, and C.W.J. Beenakker, Phys. Rev. E 
	{\bf 64}, 055203 (2001);\\
	P. Jacquod, I. Adagideli, and C.W.J. Beenakker, Europhys. Lett. 
	{\bf 61}, 729 (2003);\\
	F.M. Cucchietti, D.A.R. Dalvit, J.P. Paz, and W.H. Zurek, Phys. Rev. 
	Lett. {\bf 91}, 210403 (2003);\\	
	G. Benenti, G. Casati, and G. Veble, Phys. Rev. E {\bf 67}, 
	055202(R) (2003); \\
	G. Benenti, G. Casati, and G. Veble, Phys. Rev. E {\bf 68}, 036212 (2003);\\
	B. Eckhardt, J. Phys. A {\bf 36}, 371 (2003);\\
	T. Kottos and D. Cohen, Europhys. Lett. {\bf 61}, 431 (2003);\\
	P.G. Silvestrov, J. Tworzydlo, and C.W.J. Beenakker, Phys. Rev. E {\bf 67}, 025204(R) (2003);\\
	M. Hiller, T. Kottos, D. Cohen, and T. Geisel, Phys. Rev. Lett. {\bf 92}, 010402 (2004);\\
	T. Gorin, T. Prosen, and T. H. Seligman, New J. Phys. {\bf 6}, 20 (2004); \\
	T. Prosen and M. \v{Z}nidari\v{c}, New J. of Phys. {\bf 5}, 109 (2003);\\
	T. Prosen and M. \v{Z}nidari\v{c}, {\tt quant-ph/0401142} (2004);\\
	D.~V. Bevilaqua and E.~J. Heller, {\tt nlin.CD/0409007} (2004).

\bibitem{CERRUTI} 
	N.R. Cerruti and S. Tomsovic, Phys. Rev. Lett. {\bf 88}, 054103 (2002).

\bibitem{ANOSOV} D.V. Anosov, Proc. Steklov Inst. Math. {\bf 90}, 1 (1967).

\bibitem{BOWEN} R. Bowen, J. Diff. Eqns. {\bf 18}, 333 (1975).

\bibitem{MILLER_IVR} W.H. Miller, J. Chem. Phys., {\bf 53}, 3578 (1970).

\bibitem{MILLER_IVR_REVIEW} W.H. Miller, J. Phys. Chem. A {\bf 105}, 2942 (2001).


\bibitem{CONFERENCE} J. Van\'{\i}\v{c}ek, invited presentation, March meeting of American Physical Society, Montreal (2004).

\bibitem{UNIFIED} J. Van\'{\i}\v{c}ek, e-print: {\tt http://arXiv.org} (2004). 

\bibitem{KAPLAN} L. Kaplan, New J. Phys. {\bf 4}, 90.1 (2002). 
\bibitem{VANICEK1} J. Van\'{\i}\v{c}ek and E.J. Heller, Phys. Rev. E {\bf 64}, 026215 (2001); J. Van\'{\i}\v{c}ek and E.J. Heller, Phys. Rev. E {\bf 67}, 016211 (2003); J. Van\'{\i}\v{c}ek and D. Cohen, J. Phys. A {\bf 36}, 9591 (2003).


\end{thebibliography}
\end{document}